\documentclass[final,3p,times,twocolumn]{elsarticle}
\usepackage{amssymb}
\usepackage{amsthm}
\usepackage{amsmath}

\journal{Physica B}

\begin{document}

\begin{frontmatter}

%% Title, authors and addresses

%% use the tnoteref command within \title for footnotes;
%% use the tnotetext command for theassociated footnote;
%% use the fnref command within \author or \address for footnotes;
%% use the fntext command for theassociated footnote;
%% use the corref command within \author for corresponding author footnotes;
%% use the cortext command for theassociated footnote;
%% use the ead command for the email address,
%% and the form \ead[url] for the home page:
%% \title{Title\tnoteref{label1}}
%% \tnotetext[label1]{}
%% \author{Name\corref{cor1}\fnref{label2}}
%% \ead{email address}
%% \ead[url]{home page}
%% \fntext[label2]{}
%% \cortext[cor1]{}
%% \address{Address\fnref{label3}}
%% \fntext[label3]{}

\title{Chiral helimagnetic state in a Kondo lattice model with the Dzyaloshinskii--Moriya interaction}

%% use optional labels to link authors explicitly to addresses:
%% \author[label1,label2]{}
%% \address[label1]{}
%% \address[label2]{}

\author{Shun Okumura, Yasuyuki Kato, and Yukitoshi Motome}

\address{Department of Applied Physics, The University of Tokyo, Hongo, Bunkyo, Tokyo 113-8656, Japan} 

\begin{abstract}
%% Text of abstract

Monoaxial chiral magnets can form a noncollinear twisted spin structure called the chiral helimagnetic state. 
We study magnetic properties of such a chiral helimagnetic state, with emphasis on the effect of itinerant electrons.
Modeling a monoaxial chiral helimagnet by a one-dimensional Kondo lattice model with the Dzyaloshinskii--Moriya interaction, we perform a variational calculation to elucidate the stable spin configuration in the ground state.
We obtain a chiral helimagnetic state as a candidate for the ground state, whose helical pitch is modulated by the model parameters: the Kondo coupling, the Dzyaloshinski--Moriya interaction, and electron filling. 
\end{abstract}

\begin{keyword}
%% keywords here, in the form: keyword \sep keyword
chiral magnet, Kondo lattice model, Dzyaloshinskii--Moriya interaction, variational calculation
%% PACS codes here, in the form: \PACS code \sep code

%% MSC codes here, in the form: \MSC code \sep code
%% or \MSC[2008] code \sep code (2000 is the default)

\end{keyword}

\end{frontmatter}

%% \linenumbers

%% main text

\section{Introduction}
\label{sec:introduction}

Chirality in the lattice structure plays an important role in magnetism through the spin--orbit coupling which couples the orbital motion of electrons and the spin degree of freedom. 
It often leads to noncollinear and noncoplanar spin textures, such as a chiral helimagnetic (CHM) state~\cite{Dzyaloshinskii1964,Kataoka1981,Togawa2016} and a skyrmion crystal~\cite{Skyrme1962,Bogdanov1989,Muhlbauer2009}.
Such peculiar spin textures have attracted attention as they may result in unusual magnetoelectric phenomena, e.g., the topological Hall effect~\cite{Ohgushi2000,Lee2009,Nueubauer2009} and the spin Hall effect~\cite{Taguchi2009}.

An archetypal example of the CHM state is found in CrNb$_3$S$_6$, which is a monoaxial chiral magnet with space group of $P6_{3}22$. 
At low temperatures, the compound exhibits a CHM order at zero magnetic field, while it turns into a chiral soliton lattice (CSL) in the magnetic field applied perpendicular to the chiral axis~\cite{Moriya1982,Miyadai1983}. The CHM and CSL states were observed by using the Lorentz microscopy with a transmission electron microscope and the small-angle electron diffraction~\cite{Togawa2012}.
Theoretically, since the pioneering work by Dzyaloshinskii~\cite{Dzyaloshinskii1964,Dzyaloshinskii1965}, the CHM and CSL states have been studied for decades, whereas the most of them were limited to localized spin systems by omitting the degree of freedom for itinerant electrons~\cite{Kishine2015,Shinozaki2016,Nishikawa2016,Laliena2016}. 
Recently, the authors studied this problem by explicitly taking into account the coupling to itinerant electrons~\cite{Okumura2017}. 
Monte Carlo simulations for an extended Kondo lattice model with the Dzyaloshinskii--Moriya (DM) interaction~\cite{Dzyaloshinskii1958,Moriya1960} successfully explain a correlation between the twist of CSL and the electrical conduction.

In this paper, we report our theoretical study for the ground state of the extended Kondo lattice model whose finite-temperature properties were studied by the Monte Carlo simulations. 
Focusing on the zero-field state, we obtain the stable magnetic configuration in the ground state by a variational calculation. 
We find that the model exhibits a CHM state whose helical pitch depends on the model parameters: the Kondo coupling, the DM interaction, and electron filling.
Our results elucidate how the CHM state is stabilized by the competition between the DM interaction and an effective exchange interaction mediated by itinerant electrons.

The organization of this paper is as follows. In Section~\ref{sec:model_and_method}, we introduce a ferromagnetic Kondo lattice model with the DM interaction and the method of variational calculations. The results for the optimized magnetic structures are shown in Section~\ref{sec:result}. Section~\ref{sec:summary} is devoted to the summary.

\section{Model and method}
\label{sec:model_and_method}

Following the previous study~\cite{Okumura2017}, we consider a ferromagnetic Kondo lattice model with the DM interaction between the localized spins in one dimension.
The Hamiltonian is given by 
\begin{align}
	H =& -t\sum_{l,\nu}(c^{\dagger}_{l\nu}c^{\;}_{l+1\nu}+\mathrm{h.c.})-\mu\sum_{l,\nu}c^{\dagger}_{l\nu}c^{\;}_{l\nu}\nonumber\\
        &-J\sum_{l,\nu,\rho}c^{\dagger}_{l\nu}{\boldsymbol \sigma}_{\nu\rho}c^{\;}_{l\rho}\cdot{\mathbf S}_{l}-{\mathbf D}\cdot\sum_{l}{\mathbf S}_{l}\times{\mathbf S}_{l+1},
\label{eq:H}
\end{align}
where $c_{l\nu}(c^{\dagger}_{l\nu})$ is an annihilation (creation) operator for a $\nu$-spin electron at site $l$ on the one-dimensional chain ($\nu = \uparrow$ or $\downarrow$), $\mu$ is the chemical potential, and ${\mathbf S}_{l}=(S_l^x,S_l^y,S_l^z)$ is a three-component vector with normalized length $|{\mathbf S}_{l}|=1$. We assume the periodic boundary condition. The first term describes the kinetic energy of itinerant electrons; $t$ is a transfer integral between the nearest-neighbor sites. The third term is for the onsite coupling between the itinerant electrons and localized moments; $J$ is a positive coupling constant and $\boldsymbol{\sigma} = (\sigma^x,\sigma^y,\sigma^z)$ are the Pauli matrices. The last term represents the DM interaction with the DM vector ${\mathbf D}=D\hat{z}$, where $D>0$ and $\hat{z}$ is a unit vector along the chain direction. In this study, we focus on the case in the absence of a magnetic field. 

We study the ground state of the model in equation~(\ref{eq:H}) by a variational calculation. 
As the variational ground state, we assume a helical spin configuration represented by
\begin{align}
{\mathbf S}_{l}=(\cos{Ql},\sin{Ql},0),
\end{align}
where $Q$ is the wave number related with the helical pitch $L$ as $L = 2\pi/Q$ (we set the lattice constant as the length unit). For the spin configuration, we can calculate the energy dispersion of itinerant electrons as
\begin{align}
	\varepsilon(k) = &-2t\cos{k}\cos{\frac{Q}{2}}-\mu\nonumber\\
	&\pm\sqrt{t^2(1-\cos{2k})(1-\cos{Q})+J^2}.
\label{eq:ep}
\end{align}
Then, we can compute the total energy of the system by 
\begin{align}
	E = \sum_{-k_F \leq k < k_F}\varepsilon(k)-DN\sin{Q},
\label{eq:E}
\end{align}
where $k_F$ is the Fermi wave number [$\varepsilon(k_F)=\mu$] and $N$ is the number of sites. In the variational calculations, we optimize $E$ by varying $Q$ while tuning the chemical potential $\mu$ to set the electron filling $n$ at a particular value. The electron filling $n$ is defined by the average number of electrons per site: $n = \frac1N \sum_{-k_F \leq k < k_F}$, which varies from $0$ to $2$. The optimal value of $Q$, which we denote as $Q^*$, defines the pitch for the most stable helical spin configuration. We set the energy unit $t=1$ and take $N = 10^{4}$ in the following calculations.

\section{Result}
\label{sec:result}

\begin{figure}[h]
\centering
\includegraphics[width=\columnwidth,clip]{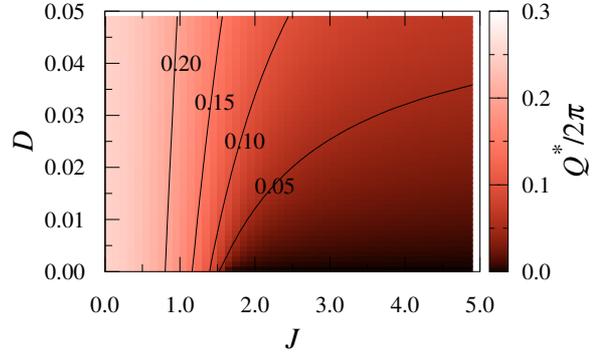}
\caption{Contour plot of the optimal wave number of the helical state, $Q^{*}/2\pi$, as a function of $J$ and $D$. $Q^{*}/2\pi$ corresponds to the inverse of the optimal helical pitch. We set the electron filling at quarter filling $n = 0.5$.}
\label{f1}
\end{figure}

Figure~\ref{f1} shows the result for the optimal wave number $Q^{*}$ as a function of $J$ and $D$ at quarter filling $n=0.5$. 
When $D=0$, an infinitesimal $J$ induces a helimagnetic order (without chirality) with the optimal wave number $Q^*=\pi/2$. 
This is due to the Ruderman--Kittel--Kasuya--Yosida interaction~\cite{Ruderman1954, Kasuya1956, Yosida1957} dictated by the twice of the Fermi wave number $2k_F = \pi/2$ at quarter filling $n=0.5$. 
While increasing $J$ at $D=0$, $Q^*$ becomes smaller (namely, the helical pitch becomes longer), and $Q^*$ vanishes for $J\gtrsim 1.6$, as shown in Fig.~\ref{f1}.
This indicates that the lowest-energy state is given by a simple ferromagnetic order without any twist for $J\gtrsim 1.6$ at $D=0$. 
The ferromagnetic state is stabilized by the effective ferromagnetic interaction mediated by the kinetic motion of itinerant electrons, called the double-exchange interaction~\cite{Zener1951, Anderson1955}. 

When $D$ is turned on, $Q^*$ increases as $D$ increases (except for $J=0$).
As shown in Fig.~\ref{f1}, however, the increase of $Q^*$ becomes slower for larger $J$. This is because the effective magnetic interaction mediated by itinerant electrons becomes stronger for a larger $J$.
In the limit of $D\rightarrow\infty$, the system prefers the helical order with $\pi/2$ rotation of spins between neighboring sites for any value of $J$, and hence, $Q^*$ converges to $\pi/2$.

\begin{figure}[h]
\begin{center}
\includegraphics[width=\columnwidth,clip]{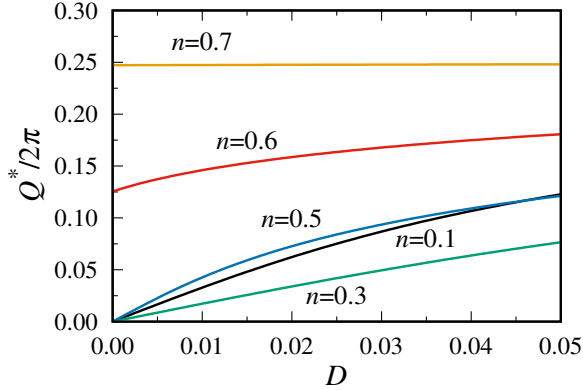}
\caption{$D$ dependence of the optimal wave number $Q^{*}/2\pi$ for several values of $n$. We take $J = 2$. 
}
\label{f2}
\end{center}
\end{figure}

We also investigate the $n$ dependence of the optimal wave number $Q^{*}$. Figure~\ref{f2} displays the values of $Q^{*}/2\pi$ as functions of $D$ at several electron fillings $n$ for $J=2$. At a low filling, for instance, at $n=0.1$, $Q^*$ is zero at $D=0$ (ferromagnetic state) and grows gradually as $D$ increases (CHM state). The growth rate is suppressed as $n$ increases because the effective ferromagnetic interaction is enhanced as the kinetic energy of itinerant electrons increases. Above $n\simeq 0.25$, 
however, the growth rate become more rapid again as $n$ increases. 
We confirm that, by Monte Carlo simulations at low temperatures, the nonmonotonic behavior of the growth rate of $Q^*$ for $D$ correlates with the stability of the ferromagnetism at $D=0$ (not shown here).
With a further increase of $n$, $Q^*$ becomes nonzero at $D=0$ above $n\simeq0.56$. This is due to a phase separation between the ferromagnetic state at $n \simeq 0.56$ and the antiferromagnetic state at half filling $n=1$~\cite{Yunoki1998}.

\section{Summary}
\label{sec:summary}

To summarize, we have studied the ground state of the one-dimensional Kondo lattice model with the DM interaction by using the variational calculation. We showed that the competition between the DM interaction and the effective magnetic interaction induced by the kinetic motion of itinerant electrons stabilizes the CHM state. 
We clarified how the helical pitch depends on the model parameters, i.e., the Kondo coupling $J$, the DM interaction $D$, and electron filling $n$. 
In the previous study~\cite{Okumura2017}, the authors performed the Monte Carlo simulations by choosing the parameters to set the helical pitch $L = 2\pi/Q^* = 10$ at zero magnetic field. 
From Fig.~\ref{f2}, we can find that this is achieved, e.g., by taking $J=2$, $D=0.035$, and $n=0.5$, which are indeed the parameters used in the previous study~\cite{Okumura2017}.

While we have focused on the magnetic state at zero magnetic field in the present study, the variational calculation can be extended to the ground state in an applied magnetic field. 
In the magnetic field applied to the chiral axis, a peculiar chiral soliton lattice (CSL) state is expected in this system, as mentioned in Sec.~\ref{sec:introduction}. 
It would be interesting to extend our variational study for investigating the optimal magnetic state while changing the magnetic field. 
In particular, it is an intriguing issue to clarify the role of itinerant electrons in the CSL state, by comparing with the previous results for the models with localized spins only.

%% The Appendices part is started with the command \appendix;
%% appendix sections are then done as normal sections
%% \appendix

%% \section{}
%% \label{}

%% If you have bibdatabase file and want bibtex to generate the
%% bibitems, please use
%%
%%  \bibliographystyle{elsarticle-num} 
%%  \bibliography{<your bibdatabase>}

%% else use the following coding to input the bibitems directly in the
%% TeX file.

\end{document}